\documentclass[9pt,twocolumn]{article}
\usepackage{times}
\usepackage[utf8x]{inputenc}
\usepackage[final]{pdfpages}
\pdfoutput=1
\usepackage{t1enc}
\usepackage{graphicx}
\usepackage{textcomp}
\usepackage[T1]{fontenc}
\usepackage{float}

\usepackage{wrapfig}

\newcommand{\Section}[1]{\vspace{-11pt}\section{\hskip -1em.~~#1}\vspace{-3pt}}

\begin{document}

\twocolumn[


\vskip-3mm 
\begin{wrapfigure}{r}{0.5\textwidth}
    \begin{flushright}
      \includegraphics{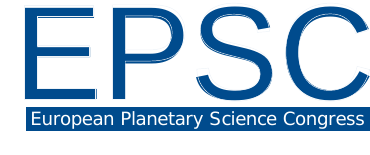}
    \end{flushright}
\end{wrapfigure}

\hspace{2 mm}  
\vskip-15mm 
\begin{flushleft}{\small 
EPSC Abstracts\\
Vol. 7 EPSC2012-784-2 2012\\
European Planetary Science Congress 2012\\
\copyright Author(s) 2012\\
}
\end{flushleft}


\vskip2.0mm

\centering{\LARGE \bf Does the innermost occurrence distribution measure tidal dissipation, reveal a flow of giant planets, or both?} 

 \vskip3mm 

 \begin{flushleft}     
{\small {\bf S. F. Taylor} (1,2) \\
(1) Unemployed, Hong Kong, (2) Participation Worldscope/Global Telescope 
Science (astrostuart@gmail.com) } 
\end{flushleft}

\vskip2mm %
]

\thispagestyle{empty}

\begin{figure}[tr!]
\end{figure}

\section*{Abstract} 

The occurrence distribution of the shortest period giant exoplanets as found by Kepler show a drop-off that is a remarkable match to the drop-off expected by taking migration due to tides in the star 
\cite{tay11}, \cite{pen12}. 
We present a comparison that can show the level of tidal dissipation (friction) as a function of the distribution of the ages of the star and planet system, with known dependencies on basic star and planet parameters. Use of this relation enables constraints to be put on the value of the tidal dissipation, constraints that will be improved as the distribution of the ages are determined. For the giant planets, this leads to an unexpectedly low value of tidal dissipation. 
This over-abundance of short period giant planets may be due to a continuing 
resupply of longer period giant planets migrating into a shorter period pileup,  disrupting the presence of smaller planets along the way. 
Perhaps the occurrence distribution of close Neptune sized planets 
will better measure the tidal friction, while the distribution of 
Jupiter sized planets reveals that giant planets are more likely
to complete a gradual migration into the star.

\Section{Introducing Infall versus $Q^{\prime}_{\ast}$ }

\begin{figure}[h]
\centerline{\includegraphics[width=\columnwidth]{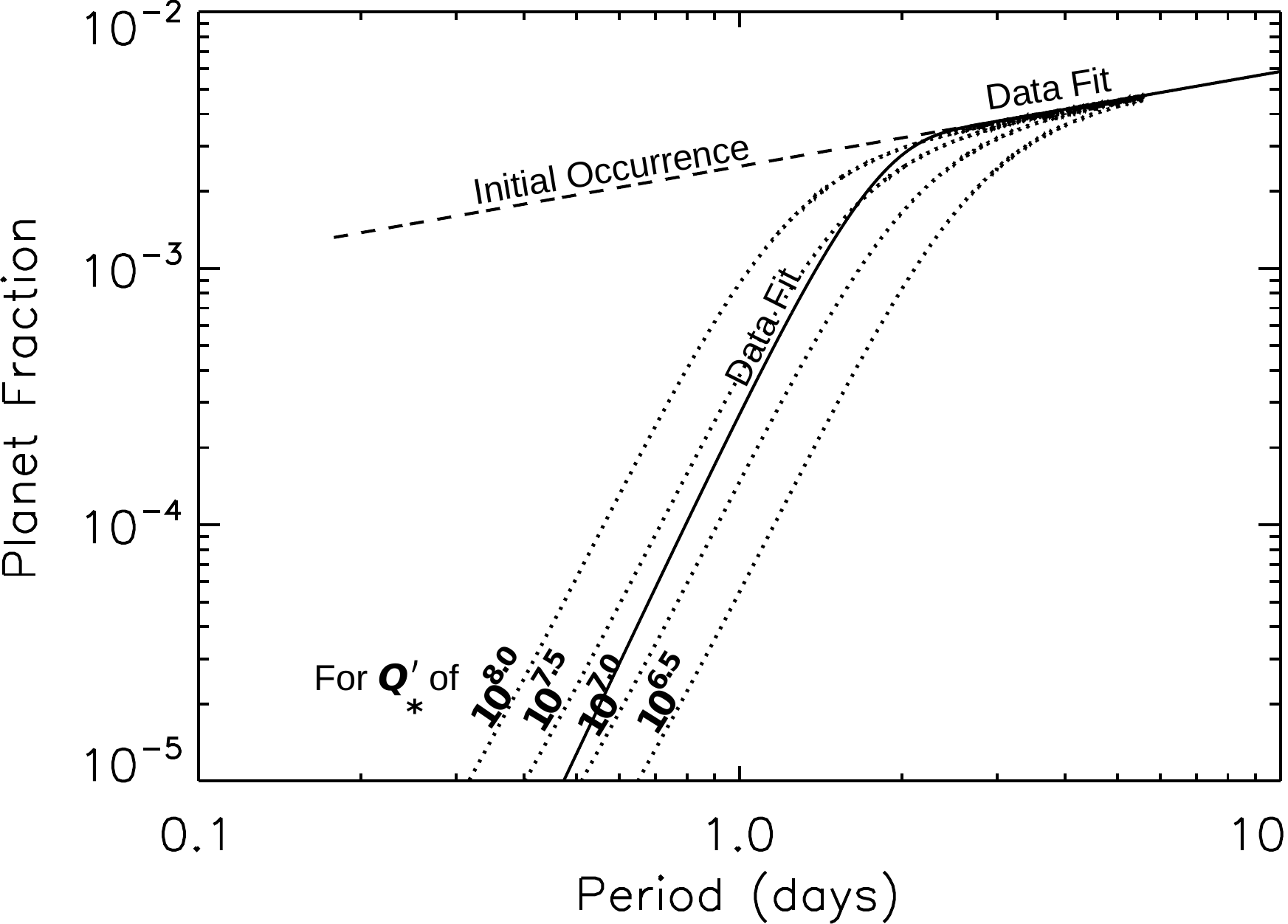}}
\caption{Comparison of a range of four calculated occurrence distribution
drop-offs (dotted lines) with the fit to 
Kepler planet candidate data (solid line) for large planets,
defined as 8 to 32 earth radii or one Jupiter mass.
found by \cite{how11} giant radii or mass planets. 
Calculated infall for four values of stellar tidal dissipation 
$Q^{\prime}_{\ast}=(10^{8.0}, 10^{7.5}, 10^{7.0}, 10^{6.5})$, is 
based on \cite{jac09} using 
solar mass stars taken to be at an age of 4.5 Gigayears, 
and using the power law of \cite{how11} without the inward dropoff
as an initial planet fraction (dashed line).
}
\label{distrijup}
\end{figure}
\begin{figure}[h]
\centerline{\includegraphics[width=\columnwidth]{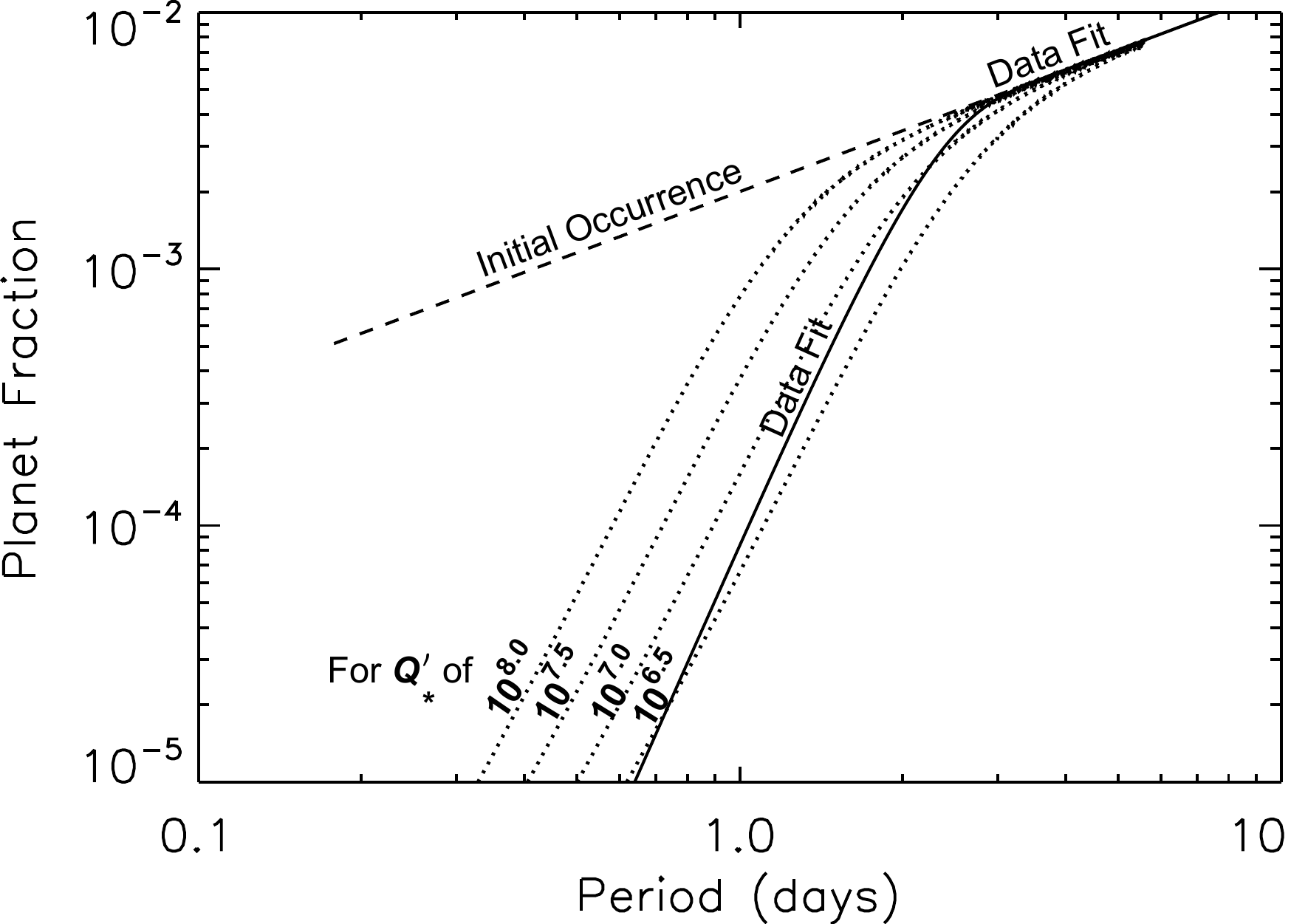}}
\caption{Similar comparison as figure~\ref{distrijup} for 
``medium planets''. The fit
to Kepler data for  4 to 8 earth radii is shown 
with calculated infall for 15 earth mass planets and solar mass stars, at an age of 4.5 Gigayears.}
\label{distrinep}
\end{figure}

The "hot Jupiter" planets, those giant planets with the shortest periods, 
have been shown to mostly be tidally migrating to their destruction by 
merger with their star, \cite{lev09}, \cite{jac09}. 
Kepler planet (candidate) occurence distributions from \cite{how11} 
show a drop-off in occurrence at short periods,
which have a power law index of 13/3, which \cite{tay11} show
is consitent with tidal migration due to tides on the star,
using the equations of \cite{jac09}.
The homogeneous Kepler statistics are sufficient to allow us to
compare calculated
drop-offs as a function of tidal dissipation (friction) $Q^{\prime}_{\ast}$
to the actual innermost occurrence distributions.
We show in figure~\ref{distrijup} and figure~\ref{distrinep} that 
these distribution give different apparent values of $Q^{\prime}_{\ast}$
for large and medium planets, which could be explained by proposing a larger
inflow of larger planets.

We find that the tidal migration increases so rapidly with smaller semi-major axis that the initial distribution matters little, so for the initial distribution we simply extend inward the power law that \cite{how11} find for the occurrence  distribution beyond the drop-off.
 Though it has long been expected that the process of planet formation would leave a drop off of planets at the shortest period, we find that the ongoing migration erases the inner-most distribution left by planet formation.



\Section{Further Work}

We are currently preparing models where the shift in the drop-off point can be interpreted in relation of tidal dissipation factor $Q^{\prime}_{\ast}$ versus realistic stellar age distribution for realistic star and planet distributions. Here we use single representative ages and star parameters for the planet masses.

\Section{Summary and Conclusions}

We compare what the tidal dissipation
$Q^{\prime}_{\ast}$ in the star 
appears to be when looking at Jupiter and Neptune mass planets. 
We find a discrepancy that could be better explained by a flow of planets
such as proposed by \cite{soc12}. This flow could be producing the ``pileup''
of hot Jupiters.
If there is a higher rate of ongoing inward migration of Jupiter mass 
than of Neptune mass planets, this could explain why we find ``too many'' 
hot Jupiters shortly before they merge with the star, without requiring
an especially low amount of tidal friction for giant planets.
This discrepancy in the apparent value of $Q^{\prime}_{\ast}$ 
is the result of the pileup of giant planets.
Compared to Neptune mass planets, 
there is a higher number of short period Jupiter mass planets 
but a lower number of longer period Jupiter mass planets. 
Kepler is now finding that the general trend of more smaller radii planets
also holds in the short period range as more smaller planets are 
detected, further emphasizing the anomalous nature of the large number
of Jupiter radii short period planets.
That the number of short period massive planets is anomalously large is made more clear by how the pattern of more planets at smaller radii in now being observed to continue into the super-earth size range at all measurable periods. 
Another sign of ongoing large planet migration is how the short period large planets are anti-correlated with the presence of other planets further out, 
but there is no such correlation for smaller short period planets. This suggests that inwardly scattered large planets make their way to the star, disrupting the orbits of smaller planets along the way, but that inwardly scattered midsize planets do not make it to the star. The resulting transient events created by the merger of planets with stars, as suggested by \cite{tay10}, \cite{met12}, and \cite{gui11}, will provide an important means of quantifying planet migration.

\subsection*{Acknowledgements}

\small

S.F.T. seeks support for continuation of this project, and also seeks participation in exoplanet confirmation projects. S.F.T. is grateful for the sacrifices of family in absence of support. These sacrifices have exceeded what should have been asked of them. We are grateful to those especially supportive members of the astronomy community who have provided discussion and assisted in the search for support.

\end{document}